\documentclass[onecolumn]{article}
\usepackage{graphicx}
\usepackage{amsmath}
\usepackage{amsfonts}
\usepackage{amssymb}

\begin{document}

\title{Time Delay in Gravitational Lensing by a Charged Black Hole of String Theory}
\author{Eduard Alexis Larra\~{n}aga R.\\Universidad Nacional de Colombia\\Observatorio Astron\'{o}mico Nacional\\}
\maketitle
\begin{abstract}
We calculate the time delay between different relativistic images formed by
the gravitational lensing produced by the
Gibbons-Maeda-Garfinkle-Horowitz-Stromiger (GMGHS) charged black hole of
heterotic string theory. Modeling the supermassive central objects of some
galaxies as GMGHS black holes, numerical values of the time delays are
estimated and compared with the correspondient Reissner-Nordstr\"{o}m black
holes . The time difference amounts to hours, thus being measurable and
permiting to distinguish between General Relativity and String Theory charged
black holes.
\end{abstract}

\section{Introduction}

\bigskip

Gravitational lensing is one of the first and more important applications of
General Relativity. The first recognized effect was the light deflection by
the sun, and after that, there was the lensing of quasars by galaxies. Now it
is an ordinary phenomenon in astronomical observations.

Because of the non-linearity of General Relativity, gravitational lensing has
been developed in the weak field approximation. However, in the last years
\ the literature is starting to look at the lensing in the strong field limit,
because of the necessity of looking for the behavior of light near massive
objects, for example near the event horizon of black holes.

The development of strong-field lensing theory was started by Virbhadra and
Ellis$^{\cite{virbhadra}}$; and more recently by Bozza$^{\cite{bozza2}}$, that
shows an analytical technique for obtaining the deflection angle in the strong
field situation and showed that the deflection angle diverge logarithmically
as light approach the photon sphere of a general class of static spherically
symmetric metrics.

\bigskip

Following the work of Bozza; A. Bhadra$^{\cite{bhadra}}$ studies the
gravitational lensing due to the Gibbons-Maeda-Garfinkle-Horowtiz-Stromiger
(GMGHS) charged black hole of heterotic string theory and compares its
estimated observables lensing quantities with those due to the
Reissner-Nordstr\"{o}m solution of General relativity. As a conclusion, Bhadra
finds that there is no significant string effect present in the angular
position and magnification of the relativistic images in the strong-gravity scenario.

\bigskip

Very recently, Bozza and Mancini$^{\cite{bozza3}}$ extended the analytical
theory of strong lensing by calculating the time delay between different
relativistic images, and shows that different types of black holes are
characterized by different time delays. Thus, this quantity can, eventually,
become available for the classification of the black holes.

\bigskip

In the present work, we follow the method of \cite{bozza3} to study the time
delay between the relativistic images produced by the GMGHS black hole
gravitational lensing. In order to get a clear idea of the string
contribution, we estimate the expected time delay for several interesting
supermassive extragalactic black holes and compare the with the expected for
the Reisnner-Nordstr\"{o}m solution.

\bigskip

The paper is structured as follows. In Sec. 2. we review the GMGHS black hole
and its strong-field lensing solution found by Bhadra. In Sec. 3. we present
the time delay computation and the estimated results for several supermassive
extragalactic black holes. These are compared with the time delays obtained
for Schwarzschild and Reissner-Nordstr\"{o}m metrics. Finally, a discussion of
the results is given in Sec. 4.

\bigskip

\section{Lensing due to the GMGHS Charged Black Hole of String Theory}

The effective action of heterotic string theory in four dimensions (in Planck
units $c=\hslash=G=1$), in the low energy sector, is given
by$^{\cite{blackside}}$:%

\begin{equation}
\mathcal{A=}\int d^{4}x\sqrt{-g}e^{-2\phi}\left(  -R+\frac{1}{12}H_{\mu v\rho
}H^{\mu v\rho}-4\left(  \nabla\phi\right)  ^{2}+F_{\mu v}F^{\mu v}\right)
\end{equation}

where $R$ is the Ricci scalar, $\phi$ is the dilaton field, $F_{\mu
v}=\partial_{\mu}A_{v}-\partial_{v}A_{\mu}$ is the Maxwell Field and $H_{\mu
v\rho}$ is a three form related to an antisymmetric tensor gauge field
$B_{\mu\nu}$ and the gauge field $A_{\mu}$ by $H=dB-A\wedge F$.

From the low energy action is clear that $e^{\phi}$ plays the role of a
coupling constant, so it governs the strength of the quantum corrections. The
complete string action includes higher order corrections $R^{2},F^{4},$ etc.
but these can be neglected when discussing black hole solutions if the size of
the hole is much larger than the Planck length.

\bigskip

Assuming that the field $H_{\mu v\rho}$ is zero, and making the conformal
transformation $g_{\mu v}^{E}=e^{-2\phi}g_{\mu v}$ to the Einstein frame, the
action can be written as:%

\begin{equation}
\mathcal{A=}\int d^{4}x\sqrt{-g_{E}}\left(  -R_{E}-2\left(  \nabla\phi\right)
^{2}+e^{-2\phi}F_{\mu v}F^{\mu v}\right)
\end{equation}

One class of solution of the above theory is the static charged black hole
configuration$^{\cite{blackside}}$, which is often called the
Gibbons-Maeda-Garfinkle-Horowtiz-Stromiger (GMGHS) black hole:%

\begin{equation}
ds^{2}=-\left(  1-\frac{2M}{r}\right)  dt^{2}+\left(  1-\frac{2M}{r}\right)
^{-1}dr^{2}+r^{2}\left(  1-\frac{Q^{2}e^{-2\phi_{o}}}{Mr}\right)  d\Omega^{2}
\label{bhstring}%
\end{equation}%

\begin{equation}
e^{-2\phi}=e^{-2\phi_{o}}\left(  1-\frac{Q^{2}e^{-2\phi_{o}}}{Mr}\right)
\end{equation}%

\begin{equation}
F_{\theta\varphi}=Q\sin\theta
\end{equation}

where $\phi_{o}$is the asymptotic constant value of the dilaton field; and $M$
and $Q$\ are the mass and electric charge \ of the black hole. The causal
structure of this space-time is identical to Schwarzschild; there is an event
horizon at $r=2M$ and a curvature singularity at $r=0$. (The vector potential
At is actually finite at $r=0$, although the invariant $F^{\mu v}F_{\mu v}$
diverges there.)

If we compare this solution with the Reissner-Nordstr\"{o}m black hole:%

\begin{equation}
ds^{2}=-\left(  1-\frac{2M}{r}+\frac{Q^{2}}{r^{2}}\right)  dt^{2}+\left(
1-\frac{2M}{r}+\frac{Q^{2}}{r^{2}}\right)  ^{-1}dr^{2}+r^{2}d\Omega^{2}%
\end{equation}

is easy to see that charged black holes of string theory are not
Reissner-Nordstr\"{o}m when $\phi=0$.

\bigskip

For the lensing study we will consider, as usual, a light ray from a source
($S$) that is deflected by a lens ($L$) and reaches the observer ($O$). We
will take the background space-time asymptotically flat. The line joining the
lens and the observer ($OL$) is the optical axis of the system. The distances
between observer and lens, lens and source and observer and source are
$D_{OL}$, $D_{LS}$ and $D_{OS}$ respectively. $\beta$ and $\theta$ are the
angular position of the source and the image respect to the optical axis; and
$\alpha$ is the deflection angle.

The position of the source and the image are related through the lens equation
obtained by Virbhadra and Ellis$^{\cite{virbhadra}}$:%

\begin{equation}
\tan\theta-\tan\beta=\frac{D_{LS}}{D_{OS}}\left[  \tan\theta+\tan\left(
\alpha-\theta\right)  \right]  \label{lens equation}%
\end{equation}

For a spherically symmetric metric,%

\begin{equation}
ds^{2}=-A\left(  r\right)  dt^{2}+B\left(  r\right)  dr^{2}+C\left(  r\right)
d\Omega^{2} \label{metric}%
\end{equation}

the deflection angle $\alpha$,is given by$^{\cite{weimberg}}$%

\begin{equation}
\alpha\left(  r_{o}\right)  =I\left(  r_{o}\right)  -\pi\label{defang}%
\end{equation}

where $r_{o}$ is the closest approach of the light ray, and%

\begin{equation}
I\left(  r_{o}\right)  =2\int_{r_{o}}^{\infty}\left[  \frac{B\left(  r\right)
}{C\left(  r\right)  }\right]  ^{\frac{1}{2}}\left[  \frac{C\left(  r\right)
A\left(  r_{o}\right)  }{C\left(  r_{o}\right)  A\left(  r\right)  }-1\right]
^{-\frac{1}{2}}dr \label{integral}%
\end{equation}

For the GMGHS black hole (\ref{bhstring}) we have:%
\begin{equation}
I\left(  x_{o}\right)  =2\int_{x_{o}}^{\infty}\frac{\frac{dx}{x}}%
{\sqrt{\left(  \frac{x}{x_{o}}\right)  ^{2}\left(  1-\frac{1}{x_{o}}\right)
\left(  1-\frac{\xi}{x}\right)  ^{2}\left(  1-\frac{\xi}{x_{o}}\right)
^{-1}-\left(  1-\frac{\xi}{x}\right)  \left(  1-\frac{1}{x}\right)  }}
\label{integralstring}%
\end{equation}

where $x_{o}=\frac{r_{O}}{2M}$; $x=\frac{r}{2M}$; $\xi=2q^{2}e^{-2\phi_{o}}$
and $q=\frac{Q}{2M}.$

The impact parameter is given by:%

\begin{equation}
u\left(  x_{o}\right)  =\frac{1}{2M}\sqrt{\frac{C\left(  x_{o}\right)
}{A\left(  x_{o}\right)  }}=x_{o}\sqrt{\frac{\left(  1-\frac{\xi}{x_{o}%
}\right)  }{\left(  1-\frac{1}{x_{o}}\right)  }}%
\end{equation}

The deflection angle increase when the closest approach $x_{o}$ decreases. For
a certain value of \ $x_{o}$ the deflection angle will be $2\pi$ and then, a
light ray will make a complete loop around the black hole. If $x_{o}$
decreases further, the light ray will give more loops, and when $x_{o}$ is
equal to the radius of the photon sphere $\left(  x_{m}\right)  $ the
deflection angle becomes infinite, and the light ray is captured by the black
hole. For the GMGHS black hole, the photon sphere is at the radius:%

\begin{equation}
x_{m}=\frac{\xi+3+\eta}{4}%
\end{equation}

with%

\begin{equation}
\eta=\sqrt{\xi^{2}-10\xi+9}%
\end{equation}

Following Bozza, the first step to solve the integral (\ref{integralstring})
in the strong field limit is to make the change of variable $z=1-\frac{x_{o}%
}{x}$, obtaining:%

\begin{equation}
I\left(  x_{o}\right)  =\int_{0}^{1}R\left(  z,x_{o}\right)  f\left(
z,x_{o}\right)  dz \label{integralstring2}%
\end{equation}

where%

\begin{equation}
R\left(  z,x_{o}\right)  =\frac{2\sqrt{1-\frac{\xi}{x_{o}}}}{1-\frac{\xi
}{x_{o}}+z\frac{\xi}{x_{o}}}%
\end{equation}%

\begin{equation}
f\left(  z,x_{o}\right)  =\left[  1-\frac{1}{x_{o}}-\left(  1-\frac{1}{x_{o}%
}+\frac{z}{x_{o}}\right)  \left(  1-z\right)  ^{2}\left(  1+\frac{z\xi}%
{x_{o}-\xi}\right)  ^{-1}\right]  ^{-\frac{1}{2}}%
\end{equation}

The function $R\left(  z,x_{o}\right)  $ is regular for all values of $z$,
while $f\left(  z,x_{o}\right)  $ diverges for $z\longrightarrow0$. Hence, the
second step is to find out the order of divergence, expanding the argument of
the square root in $f\left(  z,x_{o}\right)  $ up to the second order in $z$.
This gives:%

\begin{equation}
f\left(  z,x_{o}\right)  \sim f_{o}\left(  z,x_{o}\right)  =\frac{1}%
{\sqrt{p\left(  x_{o}\right)  z+q\left(  x_{o}\right)  z^{2}}}%
\end{equation}

where:%

\begin{equation}
p\left(  x_{o}\right)  =\frac{\eta\left(  x_{o}-2\right)  +x_{o}\left(
3-2x_{o}\right)  }{x_{o}\left(  \eta-x_{o}\right)  }%
\end{equation}%

\begin{equation}
q\left(  x_{o}\right)  =\frac{\eta^{2}-3\eta x_{o}-\left(  x_{o}-3\right)
x_{o}^{2}}{x_{o}\left(  \eta-x_{o}\right)  ^{2}}%
\end{equation}

Bozza$^{\cite{bozza2}}$ has proved that for a static spherically symmetric
space-time with the form of equation (\ref{metric}), the deflection angle
always diverge logarithmically \ for $x_{o}\longrightarrow x_{m}$; and can be
written as:%

\begin{equation}
\alpha\left(  u\right)  =-\overline{a}\log\left(  \frac{u}{u_{m}}-1\right)
+\overline{b}+O\left(  u-u_{m}\right)  \label{defang2}%
\end{equation}

All quantities with the subscript $m$ are evaluated at $x_{o}=x_{m}.$ The
coefficients of this expansion are calculated by Bhadra$^{\cite{bhadra}}$ for
the GMGHS Black hole as: :%

\begin{equation}
\overline{a}=\frac{\sqrt{3-3\xi+\eta}\left(  3-\xi+\eta\right)  }{\sqrt
{2a_{o}}} \label{a}%
\end{equation}%

\begin{equation}
\overline{b}=-\pi+\overline{a}\left\{
\begin{array}
[c]{c}%
2\left[  \log\left(  4\sqrt{a_{o}}\right)  -\log\left(  \frac{2a_{o}+a_{1}%
}{\sqrt{a_{o}}}+2\sqrt{a_{o}+a_{1}+a_{2}}\right)  \right] \\
+\log\frac{4a_{o}}{\left(  3-3\xi+\eta\right)  ^{2}\left(  -1+\xi+\eta\right)
}%
\end{array}
\right\}  \label{b}%
\end{equation}%

\begin{equation}
u_{m}=\frac{1}{2\sqrt{2}}\sqrt{\left(  9-\xi\right)  \eta+27-18\xi-\xi^{2}}
\label{u}%
\end{equation}

where:%

\begin{equation}
a_{o}=2\left(  1-\xi\right)  \left[  \xi^{2}+\xi\left(  \eta-12\right)
+9\left(  \eta+3\right)  \right]
\end{equation}%

\begin{equation}
a_{1}=4\left[  \xi^{3}+\xi^{2}\left(  \eta-15\right)  +\eta\left(
23+6\eta\right)  -3\left(  3+\eta\right)  \right]
\end{equation}%

\begin{equation}
a_{2}=24\xi^{2}-8\xi\left(  3+\eta\right)
\end{equation}

\section{Time Delay Computation}

Bozza and Mancini$^{\cite{bozza3}}$ derived the time delay betwen different
relativistic images, following an approach very similar to the one used for
the deflection angle. For an observer at infinity, the time taken for the
photon to travel from the source to the observer is:%

\begin{equation}
T=\int_{t_{o}}^{t_{f}}dt
\end{equation}

This integral can be solved following the same scheme, to obtain finally, that
the time delay between the \textit{i-th} and the \textit{j-th} relativistic
images (in Schwarzschild units) is:%

\begin{equation}
\Delta T_{i,j}^{s}=2\pi\left(  i-j\right)  \frac{\widetilde{a}}{\overline{a}%
}+2\sqrt{\frac{B_{m}}{A_{m}}}\sqrt{\frac{u_{m}}{c}}e^{\frac{\overline{b}%
}{2\overline{a}}}\left(  e^{-\frac{2\pi j\mp\gamma}{2\overline{a}}}%
-e^{-\frac{2\pi i\mp\gamma}{2\overline{a}}}\right)
\end{equation}

when the the two images are on the same side of the lens, and:%

\begin{equation}
\Delta T_{i,j}^{o}=\left[  2\pi\left(  i-j\right)  -2\gamma\right]
\frac{\widetilde{a}}{\overline{a}}+2\sqrt{\frac{B_{m}}{A_{m}}}\sqrt
{\frac{u_{m}}{c}}e^{\frac{\overline{b}}{2\overline{a}}}\left(  e^{-\frac{2\pi
j-\gamma}{2\overline{a}}}-e^{-\frac{2\pi i+\gamma}{2\overline{a}}}\right)
\end{equation}

when the two images are on opposite sides of the lens. Here $\gamma$ is th
angular separation between the source and the optical axis, as seen from the
lens. In the first case, the upper sign before $\gamma$ applies if both images
are on the same side of the source, while the lower sign if both both images
are on the other side.

\bigskip

For spherically symmetric metrics, we have:%

\begin{equation}
\frac{\widetilde{a}}{\overline{a}}=\sqrt{\frac{C_{m}}{A_{m}}}%
\end{equation}%

\begin{equation}
c=\frac{C_{m}^{\prime\prime}A_{m}-C_{m}A_{m}^{\prime\prime}}{4\sqrt{A_{m}%
^{3}C_{m}}}%
\end{equation}

When the source is almost aligned with the lens, we have $\gamma\sim
D_{OL}^{-1}\ll2\pi$; and therefore, in this case, we have for $i\neq j$.:%

\begin{equation}
\Delta T_{i,j}^{s}\simeq\Delta T_{i,j}^{o}%
\end{equation}

\subsection{Time Delay in Supermassive GMGHS Black Hole Lensing}

By applying this scheme for the GMGHS black hole and assuming that the source
is almost aligned with the lens, we obtain the time delay between the second
and the first relativistic images as:%

\begin{equation}
\Delta T_{2,1}=2\pi\frac{\widetilde{a}}{\overline{a}}+2\sqrt{\frac{B_{m}%
}{A_{m}}}\sqrt{\frac{u_{m}}{c}}e^{\frac{\overline{b}}{2\overline{a}}}\left(
e^{-\frac{\pi}{\overline{a}}}-e^{-\frac{2\pi}{\overline{a}}}\right)
\end{equation}

where the coefficients are:%

\begin{equation}
A_{m}=\left(  1-\frac{1}{x_{m}}\right)  =B_{m}^{-1}%
\end{equation}%

\begin{equation}
\frac{\widetilde{a}}{\overline{a}}=\sqrt{\frac{\left(  x_{m}^{2}-\xi
x_{m}\right)  }{\left(  1-\frac{1}{x_{m}}\right)  }}%
\end{equation}%

\begin{equation}
c=\frac{2\left(  1-\frac{1}{x_{m}}\right)  +\frac{1}{x_{m}}\left(  1-\frac
{\xi}{x_{m}}\right)  }{4\sqrt{\left(  1-\frac{1}{x_{m}}\right)  ^{3}\left(
x_{m}^{2}-\xi x_{m}\right)  }}%
\end{equation}

And $\overline{a},\overline{b}$ and $u_{m}$ given by (\ref{a}),(\ref{b}), and
(\ref{u}) respectively. Thus, the time delay for the GMGHS black hole depends
on its mass and electric charge.

\bigskip

Now we will give some realistic values by modelling some supermassive black
holes as GMGHS. Taking the case $q=0$ (i.e. no electric charge) we obtain the
same time delays reported by Bozza and \ Macini$^{\cite{bozza3}}$ for the
Scharzschild metric:

\bigskip

\begin{center}%
\begin{tabular}
[c]{|c|c|c|}\hline
Galaxy & Mass ($M_{\odot}$) & $\Delta T_{2,1}$ \ \ GMGHS $\left(  q=0\right)
$\\\hline
NGC4486 (M87) & $3.3\times10^{9}$ & $149.3$ h.\\\hline
NGC3115 & $2.0\times10^{9}$ & $90.5$ h.\\\hline
NGC4374 (M84) & $1.4\times10^{9}$ & $63.3$ h.\\\hline
NGC4594 (M104) & $1.0\times10^{9}$ & $45.2$ h.\\\hline
NGC4486B (M104) & $5.7\times10^{8}$ & $25.8$ h.\\\hline
NGC4261 & $4.5\times10^{8}$ & $20.4$ h.\\\hline
NGC4342 (IC3256) & $3.9\times10^{8}$ & $17.6$ h.\\\hline
NGC7052 & $3.3\times10^{8}$ & $14.9$ h.\\\hline
NGC3377 & $1.8\times10^{8}$ & $8.1$ h\\\hline
\end{tabular}

\bigskip Table I. Time delay for GMGHS Black Hole.without electric charge
\end{center}

This result just shows that the GMGHS black hole without electric charge
reduces to the Schwarzschild metric.

Now, if we put some electric charge ($q\neq0$) the estimated time delay for
the same black holes differ from the obtained by Bozza and Mancini for the
Reissner-Nordstr\"{o}m metric:

\bigskip%

\begin{tabular}
[c]{|c|c|c|c|c|c|c|c|c|}\hline
& \multicolumn{4}{|c|}{Reissner-Nordstr\"{o}m $\Delta T_{2,1}$ \ (hours.)} &
\multicolumn{4}{|c|}{GMGHS $\Delta T_{2,1}$\ \ (hours.)}\\\cline{2-9}%
Galaxy & $q=0.1$ & $q=0.2$ & $q=0.3$ & $q=0.4$ & $q=0.1$ & $q=0.2$ & $q=0.3$ &
$q=0.4$\\\hline
NGC4486 & $148.3$ & $145.4$ & $140.1$ & $131.7$ & $148.4$ & $145.\,5$ &
$140.5$ & $133.1$\\\hline
NGC3115 & $89.9$ & $88.1$ & $84.9$ & $79.8$ & $90.0$ & $88.\,2$ & $85.2$ &
$80.7$\\\hline
NGC4374 & $62.9$ & $61.7$ & $59.4$ & $55.9$ & $63.0$ & $61.7$ & $59.6$ &
$56.5$\\\hline
NGC4594 & $44.9$ & $44.0$ & $42.4$ & $39.9$ & $45.0$ & $44.1$ & $42.\,6$ &
$40.3$\\\hline
NGC4486B & $26.6$ & $25.1$ & $24.2$ & $22.7$ & $25.6$ & $25.1$ & $24.3$ &
$23.0$\\\hline
NGC4261 & $20.2$ & $19.8$ & $19.1$ & $18.0$ & $20.2$ & $19.\,8$ & $19.2$ &
$18.1$\\\hline
NGC4342 & $17.5$ & $17.2$ & $16.6$ & $15.6$ & $17.5$ & $17.\,2$ & $16.6$ &
$15.7$\\\hline
NGC7052 & $14.8$ & $14.5$ & $14.0$ & $13.2$ & $14.8$ & $14.6$ & $14.1$ &
$13.3$\\\hline
NGC3377 & $8.1$ & $7.9$ & $7.6$ & $7.2$ & $8.1$ & $7.9$ & $7.7$ &
$7.3$\\\hline
\end{tabular}

\begin{center}
\bigskip\ \ \ \ \ \ \ \ \ \ Table II. Time delays for GMGHS and
Reissner-Nordstr\"{o}m Black Holes.

\bigskip
\end{center}

\section{Conclusion}

Since the charged black holes of General Relativity and String Theory are
qualitatively different, it is expected that there will be some differences in
the observational features of each of them, particularly in the strong field limit.

Here, we have used the analytical model of Bozza to obtain the time delay in
the strong field scenario for the GMGHS black hole. Modeling some supermassive
objects at the center of galaxies as charged string theory black holes, we
have seen that there are some differences in the numerical estimates of the
time delay between the second and the first relativistic images, when compared
with the obtained in the Reissner-Nordstr\"{o}m case by Bozza and Mancini. \ 

As can be seen from Table II, the string effects in the time delay are more
evident for great masses and great electric charges. The differences in the
estimates are of the order of hours, so they can be easy measured, in
principle; providing a possible method for distinguish between General
Relativty and String Theory charged black holes.


\begin{thebibliography}{9}
\bibitem{bhadra}A. Bhadra, Phys. Rev. D \textbf{67}, 103009 (2003)

\bibitem{blackside}G.T. Horowitz, Proceedings of the Trieste Spring School on
String Theory an Quantum Gravity. 1992

\bibitem{virbhadra}K.S. Virbhadra and George F.R. Ellis, Phys.Rev.D
\textbf{62}, 084003 (2000)

\bibitem{weimberg}S. Weinberg, \textit{Gravitation and Cosmology: Principles
and Applications of the General Theory of Relativity}. Jhon\ Wiley, NY. 1972

\bibitem{bozza}V. Bozza, S. Capozziello, G. Ioavane and G. Scarpetta, Gen.
Rel. Grav. \textbf{33}, 135 (2001)

\bibitem{bozza2}V. Bozza, Phys. Rev. D. \textbf{66}, 103001 (2002)

\bibitem{bozza3}V. Bozza and L. Mancini.\ gr-qc/0305007
\end{thebibliography}
\end{document}